\documentclass[12pt]{article}

\usepackage{graphics}
\usepackage{amssymb}
\usepackage[driver]{graphicx}

\newcommand{\QED}{\mbox{\rule[-1.5pt]{6pt}{10pt}}}
\newcommand{\C}{\mathbb{C}}
\newcommand{\N}{\mathbb{N}}
\newcommand{\R}{\mathbb{R}}

\newcommand{\e}{{\rm e}}
\newcommand{\im}{{\rm Im\,}}
\newcommand{\D}{{\rm d}}
\newcommand{\DD}{{\cal D}}
\newcommand{\RR}{{\cal R}}

\newcommand{\OO}{{\cal O}}

\newcommand{\SSS}{{\cal S}}
\newcommand{\ageq}{\stackrel{>}{\scriptstyle\sim}}

\newtheorem{claim}{Claim}[section]
\newtheorem{theorem}[claim]{Theorem}
\newtheorem{proposition}[claim]{Proposition}
\newtheorem{lemma}[claim]{Lemma}
\newtheorem{corollary}[claim]{Corollary}
\newtheorem{example}[claim]{Example}
\newtheorem{remark}[claim]{Remark}

\begin{document}

\title{Leaky Quantum Graphs: Approximations \\
by Point Interaction Hamiltonians}
\date{}
\author{P.~Exner,$^{1,2}$ K.~N\v{e}mcov\'{a}$^{1,3}$}
\maketitle
\begin{quote}
{\small \em 1 Nuclear Physics Institute, Academy of Sciences,
25068 \v Re\v z \\ \phantom{a)}near Prague, Czechia \\ 2 Doppler
Institute, Czech Technical University, B\v rehov{\'a} 7,
\\ \phantom{a)}11519 Prague, Czechia \\ 3 Institute of Theoretical Physics,
FMP, Charles University, \\ \phantom{a)}V Hole\v{s}ovi\v{c}k\'ach
2, 18000 Prague, Czechia}
\end{quote}

\begin{abstract}
\noindent We prove an approximation result showing how operators
of the type $-\Delta -\gamma \delta (x-\Gamma)$ in
$L^2(\mathbb{R}^2)$, where $\Gamma$ is a graph, can be modeled in
the strong resolvent sense by point-interaction Hamiltonians with
an appropriate arrangement of the $\delta$ potentials. The result
is illustrated on finding the spectral properties in cases when
$\Gamma$ is a ring or a star. Furthermore, we use this method to
indicate that scattering on an infinite curve $\Gamma$ which is
locally close to a loop shape or has multiple bends may exhibit
resonances due to quantum tunneling or repeated reflections.
\end{abstract}



\section{Introduction}

The main aim of this paper is to discuss a limiting relation
between two classes of generalized Schr\"{o}dinger operators in
$L^2(\mathbb{R}^2)$. One of them are measure-type perturbations of
the Laplacian, in particular, we will be interested in potentials
which are a negative multiple of the Dirac measure supported by a
finite graph $\Gamma\subset \mathbb{R}^2$, in other words, they
are formally given by the expression
\begin{equation} \label{formal}
-\Delta -\gamma \delta (x-\Gamma )
\end{equation}
with some $\gamma>0$. We are going to show that such operators can
be approximated in the strong resolvent sense by families of
point-interaction Hamiltonians \cite{AGHH} with two-dimensional
$\delta$ potentials suitably arranged.

Stated in this way the problem is an interesting mathematical
question; the indicated result will represent an extension the
approximation theorem derived in \cite{BFT} to the two-dimensional
situation, which is not straightforward because properties of the
$\delta$ potentials are dimension-dependent. In addition, we have
a strong physical motivation coming from the fact that
Hamiltonians of the type (\ref{formal}) are good models of various
graph-type nanostructures which in distinction to the usual
description \cite{KS} take quantum tunneling into account -- see
\cite{Ex1} for a bibliography to this problem.

To investigate such models one has to find spectral properties of
the operator (\ref{formal}) for various shapes of $\Gamma$. This
is in general a complicated task even if the original PDE problems
is reformulated by means of the generalized Birman-Schwinger
principle \cite{BEKS} into solution of an appropriate integral
equation. On the other hand, spectral problem for
point-interaction Hamiltonians is reduced in a standard way to
solution of an algebraic equation with coefficients containing
values of the free Green function. Hence an approximation of the
mentioned type would be of practical importance; our second aim is
to illustrate this aspect of the problem on examples.

Let us describe briefly contents of the paper. In the next section
we collect the needed preliminaries stated in a way suitable for
the further argument. In Section~\ref{main result} we formulate
and prove our main result, which is the approximation indicated
above. In the following two sections we discuss examples of two
simple graph classes. The first are graphs of the form of a ring,
full or open. In this case the spectrum of the operator
(\ref{formal}) can be found by mode matching \cite{ET} which
enables us to compare the approximation with the ``exact'' result,
in particular, to assess the rate of its convergence. In contrast,
in Section~\ref{stargraph} we discuss star-shaped graphs. Here an
alternative method for numerical solution of the spectral problem
is missing, however, we can derive several conclusions for (an
infinite) star analytically and show how does the approximation
results conform with them. Computations of this type were
performed already in \cite{EN} but the number of points used there
was too small to provide a reasonable approximation.

In the last section we address another question which is more
difficult and no analytical results are available presently. It
concerns the scattering problem on infinite leaky graphs with
asymptotically straight ``leads''. On a heuristic level, it is
natural to expect that the states from the negative part of the
continuous spectrum can propagate being transversally confined to
the graph edges, hence a nontrivial geometry should yield an
$N\times N$ on-shell scattering matrix, where $N$ is the number of
the leads. This remains to be proved, however, and even more
difficult it will be to compute the S-matrix mentioned above in
terms of the graph geometry.

For simplicity we will restrict ourselves to the simplest case
when $\Gamma$ is an infinite asymptotically straight curve, i.e.
$N=2$, without self-intersections. One natural conjecture is that
if the distance between the curve points is small somewhere so
that $\Gamma$ is close to a loop shape locally, the system can
exhibit resonances due to quantum tunneling. On the other hand,
simple bends are unlikely to produce distinguished resonances; the
reason is that the transverse $\delta$ coupling in (\ref{formal})
has a single bound state, hence there is no analog here to higher
thresholds which give rise to resonances in bent hard-wall tubes
\cite{DEM}. To support these conjectures, we have let us inspire
by the so-called $L^2$ approach to resonances, which was in the
one-dimensional case set on a rigorous ground in \cite{HM}.
Specifically, we study the spectrum of a finite segment of $\Gamma$
as a function of the cut-off position. In the first case we find
in this dependence avoided eigenvalue crossings, the more narrow
the smaller the gap to tunnel is, while no such effect is seen in
the second case.

On the other hand, a tunneling between arcwise distant points of
the graphs is not the only source of resonances. To illustrate
this claim, we analyze in our last example a Z-shaped graph with
two sharp bends separated by a line segment. We find again avoided
crossings the widths of which vary widely as functions of the
bending angle. This provides an indirect but convincing indication
that the above described on-shell S-matrix for a bent curve in the
shape of a broken line is nontrivial.


\setcounter{equation}{0}
\section{Preliminaries} \label{prelim}

\subsection{Generalized Schr\"odinger operators}

We start with the definition of Schr\"odinger operators of the
form $-\Delta - \gamma m$ in $L^2(\R^2)$, where $m$ is a finite
positive measure on the Borel $\sigma$-algebra of $\Gamma$, which
is assumed to be a non-empty closed subset of $\R^2$ and also the
support of the measure $m$. Furthermore, $\gamma$ is a bounded and
continuous function, acting from $\Gamma$ to $\R_+$. We suppose that
the measure $m$ belongs to the generalized Kato class, in
two-dimensional case it means that the following condition holds,
   \begin{equation}\label{Kato}
\lim_{\varepsilon \to 0} \sup_{x \in \R^2} \int_{B(x,\varepsilon)}
|\log (|x-y|)| m(dy) = 0\,,
   \end{equation}
where $B(x,\varepsilon)$ denotes the circle of radius $\varepsilon$
centered at $x$.

Due to the fact that the measure $m$ belongs to Kato class, for
each $a>0$ there exists $b \in \R$ such that any $\psi$ from the
Schwartz space $\SSS(\R^2)$ satisfies the inequality
  \begin{equation}\label{m estimate}
\int_{\R^2} |\psi(x)|^2 m(dx) \leq a \int_{\R^2} |\nabla
\psi(x)|^2 dx + b \int_{\R^2} |\psi(x)|^2 dx\,,
  \end{equation}
as it was proven in the paper \cite{SV}. Since $\SSS(\R^2)$ is
dense in $H^1(\R^2)$ we can define a bounded linear transformation
  \begin{eqnarray*}
I_m : \; H^1(\R^2) &\mapsto& L^2(m) \\ I_m \psi&=&\psi\,, \qquad
\forall \psi \in \SSS(\R^2)\,.
  \end{eqnarray*}
Using this transformation, the inequality (\ref{m estimate}) can
be extended to the whole $H^1(\R^2)$ with the function $\psi$ on
l.h.s. replaced by $I_m \psi$. By employing the KLMN theorem, see
\cite[Thm.~X.17]{RS}, we conclude that the quadratic form
$q_{\gamma m}$ given by
  \begin{eqnarray} \label{form}
D(q_{\gamma m}) &:=& H^1(\R^2)\,, \\ q_{\gamma m} (\psi,\phi) &:=&
\int_{\R^2} \nabla \bar{\psi}(x) \nabla \phi(x) dx - \int_{\R^2}
I_m\bar{\psi}(x) I_m \phi(x) \gamma(x) m(dx)\,, \nonumber
  \end{eqnarray}
is lower semi-bounded and closed in $L^2(\R^2)$. Hence there
exists a unique self-adjoint operator $H_{\gamma m}$ associated
with the form $q_{\gamma m}$.

The described definition applies to more general class of measures
than we need here. If $\Gamma$ is a graph consisting of a locally
finite number of smooth edges meeting at nonzero angles, i.e.
having no cusps, there is another way to define $H_{\gamma m}$,
namely via boundary conditions imposed on the wavefunctions. To
this aim, consider first the operator $\dot H_{\gamma m}$ acting
as
$$ \left(\dot H_{\gamma m} \right)(x) = -(\Delta\psi)(x)\,,
\quad x\in \R^2\setminus\Gamma\,, $$
for any $\psi$ of the domain consisting of functions which belong
to $H^2(\R^2\setminus\Gamma)$, are continuous at $\Gamma$ with the
normal derivatives having there a jump,
\begin{equation} \label{jump}
{\partial\psi\over\partial n_+}(x) - {\partial\psi\over\partial n_-}
(x) = -\gamma\psi(x)\,, \quad x\in\Gamma\,.
\end{equation}
Then it is straightforward to check that $\dot H_{\gamma m}$ is
e.s.a. and by Green's formula it reproduces the form $q_{\gamma
m}$ on its core, so its closure may be identified with $H_{\gamma
m}$ defined above.

An important tool to analyze spectra of such operators is the
generalized Birman-Schwinger method. If $k^2$ belongs to the
resolvent set of $H_{\gamma m}$ we put $R^k_{\gamma m}:=
(H_{\gamma m}-k^2)^{-1}$. The free resolvent $R^k_0$ is defined
for $\im k>0$ as an integral operator with the kernel
\begin{equation} \label{freeG}
G_k(x\!-\!y) = {i\over 4}\, H_0^{(1)} (k|x\!-\!y|)\,.
\end{equation}
Now we shall use $R^k_0$ to define three other operators. For the
sake of generality, suppose that $\mu, \nu$ are positive Radon
measures on $\R^2$ with $\mu(x)= \nu(x) =0$ for any $x\in\R^2$. In
our case they will be the measure $m$ on $\Gamma$ and the
Lebesgue measure $\D x$ on $\R^2$ in different combinations. By
$R^k_{\nu,\mu}$ we denote the integral operator from
$L^2(\mu)$ to $L^2(\nu)$ with the kernel $G_k$, i.e.
$$ R^k_{\nu,\mu} \phi = G_k \ast \phi\mu $$
holds $\nu$-a.e. for all $\phi\in D(R^k_{\nu,\mu}) \subset
L^2(\mu)$.

With this notation one can express the resolvent
$R^k_{\gamma m}$ as follows \cite{BEKS}:
\begin{theorem} \label{gen krein}
Let $\im k>0$. Suppose that $I-\gamma R^k_{m,m}$ is
invertible and the operator
$$ R^k := R_0^k + \gamma R^k_{\D x,m} [I-\gamma R^k_{m,m}]^{-1}
R^k_{m,\D x} $$
from $L^2(\R^2)$ to $L^2(\R^2)$ is everywhere defined. Then $k^2$
belongs to $\rho(H_{\gamma m})$ and $(H_{\gamma m} -k^2)^{-1}=
R^k$.
\end{theorem}

The invertibility hypothesis is satisfied for all sufficiently
large negative $k^2$ because for such a $k^2$ the operator norm of
$\gamma R_{m,m}(z)$ acting in $L^2(m)$ is less than 1, see again
\cite{BEKS}, and similar result can be proven for the operator
norm in $L^\infty(m)$ following \cite{BFT}. Thus from now on we
consider $k^2<0$ such that both these norms are less than one.

For later considerations it is useful we rewrite operator
$H_{\gamma m}=-\Delta- \gamma m$ in the form $-\Delta - {1 \over
\alpha} \mu$, where we have introduced
  \begin{equation}\label{def alpha}
\mu = {\gamma m \over \int \gamma m}\,, \qquad \alpha = {1
\over \int \gamma m}\,. 
  \end{equation}
Since function $\gamma$ acquires only non-negative values and $m$ is
positive measure, $\alpha$ is a positive number. The resolvent reads
  \begin{equation}\label{sing krein}
R^k_{\gamma m}=R^k_0 + R^k_{\D x,\mu} \left( 1- {1\over \alpha}
R^k_{\mu,\mu} \right)^{-1} {1\over \alpha} R_{\mu,\D x}^k\,.
  \end{equation}
It can be rewritten alternatively as
  \begin{equation}\label{sing krein alt}
(H_{\gamma m}-z)^{-1}\psi = R^k_0 \psi + R_{\D x,\mu}^k \sigma,
  \end{equation}
where $\psi \in L^2(\R^2)$ and $\sigma \in L^2(\mu)$ represents
the unique solution to the equation
  \begin{equation}\label{eq sigma}
\alpha \sigma - R_{\mu,\mu}^k \sigma = R_{\mu,\D x}^k \psi \qquad
\mu-a.e.
  \end{equation}
Since $R^k_0 \psi$ belongs to $H^2(\R^2)$, using Sobolev's embedding
theorem we conclude that it has a bounded and continuous version,
and therefore $\sigma$ also has a representative which is bounded
and continuous on the set $S_\gamma:=\{x\in\Gamma:\: \gamma(x)\ne
0\}$.


\subsection{Schr\"odinger operators with point interactions}

Consider a discrete and finite subset $Y \subset \Gamma$ and the
positive constant $\alpha$ defined above. As it is well known we
cannot regard the operator $H_{Y,\alpha}$ with the interaction
supported by $Y$ as before, i.e. as `$-\Delta$ + measure'. Instead,
we define this operator via its domain: each function $\psi \in
D(H_{Y,\alpha})$ behaves in the vicinity of a point $a \in Y$ as
follows
  \begin{equation}
\psi (x) = - {1 \over 2\pi} \log|x-a| \, L_0(\psi,a) + L_1(\psi,a)
+ \OO(|x-a|),
  \end{equation}
where the generalized boundary values $L_0(\psi,a)$ and
$L_1(\psi,a)$ satisfy
  \begin{equation} \label{bc}
L_1(\psi,a) + 2\pi |Y|\alpha\, L_0(\psi,a) = 0
  \end{equation}
for any $a \in Y$ with $|Y|:= \sharp(Y)$; for a justification of
this definition and further properties of the operators
$H_{Y,\alpha}$ see, e.g., \cite[Chap.~II.4]{AGHH}.

The form (\ref{bc}) which we have chosen is adapted for the
indicated use of these operator in the approximation. In
particular, the coupling parameter $|Y|\alpha$ depends linearly on
the cardinality of the set $Y$. However, it is important to stress
that our problem differs substantially form its three-dimensional
analogue considered in \cite{BFT} where no sign of $\alpha$ played
a preferred role. In the two-dimensional setting the coupling
parameter tends to $\pm\infty$ in the limit of weak and strong
coupling, respectively. Consequently, we will be able to find an
approximation for operators (\ref{formal}) with \emph{attractive
interactions} only.

The Krein's formula for the resolvent $(H_{Y,\alpha}-z)^{-1}$ is
the basic ingredient in the spectral analysis of the
point-interaction Hamiltonians.
\begin{theorem}
Let $k^2<0$ and $\Lambda_{Y,\alpha}(k^2)$ be the matrix
$|Y|\times|Y|$ given by
  \begin{eqnarray}
\Lambda_{Y,\alpha}(k^2;x,y) &=& {1\over 2\pi} \left[ 2\pi |Y|
\alpha + \log \left( {i k\over 2} \right)+C_E \right]
\delta_{xy} \nonumber \\ && - G_k(x-y)(1-\delta_{xy})\,,
  \end{eqnarray}
where $C_E$ is the Euler constant. Suppose that this matrix is
invertible. Then $k^2 \in \rho(H_{Y,\alpha})$ and we have
  \begin{eqnarray}\label{point krein}
(H_{Y,\alpha}-k^2)^{-1}\psi \,(x) = R^k_0 \psi \, (x) &+&
\sum_{y,y' \in Y} \left[ \Lambda_{Y,\alpha}(k^2)\right]^{-1}
(y,y')\, G_k (x-y) \nonumber
\\ && \phantom{AAAAA} \times R_0^k \psi \,(y')
  \end{eqnarray}
for any $\psi \in L^2(\R^2)$.
\end{theorem}
One can see easily that the matrix $\Lambda_{Y,\alpha}(k^2)$ is
invertible for sufficiently large negative $z=-\kappa^2$. Indeed,
the diagonal part is dominated by the term ${1\over 2\pi} \log
\kappa$, while the non-diagonal elements vanish as $\kappa \to
\infty$, see the asymptotic formula \cite[9.2.7]{AS} for the
Hankel function $H_0^{(1)}$. In view of our special choice of the
coupling, an alternative way how to make the matrix
$\Lambda_{Y,\alpha}$ invertible is to take a sufficiently large
set $Y$.
\begin{lemma}\label{lemma: lambda}
Let $\im k >0$ and $(Y_n)_{n \in \N}$ be a sequence of non-empty
finite subsets of $S_\gamma$ such that $|Y_n| \to \infty$ as $n
\to \infty$ and the following inequality holds
  \begin{equation}\label{Schur test}
\sup_{n \in \N} {1 \over |Y_n|} \sup_{x \in Y_n} \sum_{y \in Y_n
\setminus \{x\} } G_k (x-y) < \alpha.
  \end{equation}
Then there exists a positive $C$ and $n_0 \in \N$ such that the
matrix $\Lambda_{Y_n,\alpha}(k^2)$ is invertible and
  \begin{equation}
\Big\| \left( {1 \over |Y_n|}\, \Lambda_{Y_n,\alpha}(k^2) \right)^{-1}
\Big\|_{2,2} < C
  \end{equation}
holds for all $n \geq n_0$. Here $\|\cdot\|_{p,q}$ means the norm
of the map from $\ell^p$ to $\ell^q$; the specification is
superfluous here but it will be useful in the following.
\end{lemma}
{\sl Proof:} Let us first decompose the matrix $1/|Y_n| \,
\Lambda_{Y_n, \alpha}(k^2)$ into the diagonal and non-diagonal
parts, $D_n$ and $R_n$, respectively. For $n$ being large enough
the diagonal matrix $D_n$ is invertible and its operator norm in
$(\C^{|Y_n|},\|\cdot\|_2)$ converges to $\alpha$ as $n \to
\infty$. Due to the strict inequality in the hypothesis there is
an $a<\alpha$ such that the inequality (\ref{Schur test}) holds
with $\alpha$ replaced by $a$. Then the Schur-Holmgren bound
\cite[App.~C]{AGHH} yields $\| R_n \|_{2,2} \le a <\alpha$,
which in turn implies that the matrix sum $D_n +R_n$ is invertible
for a sufficiently large $n$. \QED \vspace{.5ex}

In analogy with the expression (\ref{sing krein alt}) it is
possible to rewrite Krein's formula (\ref{point krein}) in the
form
  \begin{equation}\label{point krein alt}
(H_{Y,\alpha}-k^2)^{-1}\psi \,(x) = R_0^k \psi \, (x) + \sum_{y
\in Y} G_k (x-y)\, q_y\,,
  \end{equation}
where $q_y, y\in Y_n$ solve the following system of equations,
  \begin{equation}\label{eq q}
{1\over 2\pi} \left[ 2\pi |Y| \alpha + \log \left( {i k\over 2}
\right)+C_E \right] q_y - \sum_{y' \in Y, y'\neq y} G_k (y-y')
q_{y'} = R_0^k \psi \, (y)\,,
  \end{equation}
for all $y \in Y_n$.


\setcounter{equation}{0}
\section{Approximation by Schr\"odinger Operators with Point Interactions}
\label{main result}

With the above preliminaries we can proceed to the main goal -- we
will prove that for a chosen generalized Schr\"odinger operator
with an attractive interaction one can find an approximating
sequence of point-potential Schr\"odinger operators under
requirements which will be specified below.

The assumption about positions of the point potentials is obvious
-- loosely speaking, as the sites of potentials are getting denser
in the set $\Gamma$, they must copy the measure $\mu$. Then we
have to specify the coupling-constant behaviour in the
approximating operators. We have already mentioned that in the
analogous situation in dimension three the coupling constants
scale by \cite{BFT} linearly with the number $|Y|$ of point
potentials, and suggested the same behaviour here. This requires
an explanation, because it is well known that the coupling
constants are manifested differently in dimension three and two;
just consider a pair of point potentials and let their distance
vary.

To get a hint that the scaling behavior for the approximation
remains nevertheless the same, consider an infinite straight
polymer as in \cite[III.4]{AGHH}, denote the coupling constant by
$\alpha$ and the period by $l_0 n^{-1}$. The threshold of the
continuous spectrum is given as the unique solution $E=-\kappa^2$
to the implicit equation
  \begin{equation}\label{polymer}
\alpha = {n \over 2 l_0 \kappa} - {1 \over 2\pi} \log
{2\pi n \over l_0} +\lim_{M \to \infty} \sum_{m=1}^M
\left( {1 \over \sqrt{(2\pi m)^2+(\kappa l_0 /n)^2}} -
{1 \over 2\pi m} \right).
  \end{equation}
Now let the number $n$ increase. If we want to keep the solution
$\kappa$ preserved as $n\to\infty$, then $\alpha$ must grow
linearly with $n$; recall that for $\alpha>0$ this means that the
individual point interactions are becoming \emph{weaker}. This
motivates the choice of the coupling constants in the form $|Y|
\alpha$ which we made in the boundary condition (\ref{bc}).

Now we can prove the announced approximation result.
\begin{theorem}
Let $\Gamma$ be a closed and non-empty subset of $\R^2$ and let
$m$ be a finite positive measure on Borel $\sigma$-algebra of
$\Gamma$ with $\mathrm{supp}\,m=\Gamma$, which belongs to the Kato
class. Let $\gamma: \Gamma \to \R_+$ be a nontrivial bounded
continuous function. Choose $k$ with $\im k>0$ such that the
equation (\ref{eq sigma}) has a unique solution $\sigma$ which has
a bounded and continuous version on $S_\gamma:=\{x\in\Gamma:\:
\gamma(x)\ne 0\}$. Finally, suppose that there exists a sequence
$( Y_n )_{n=1}^\infty$ of non-empty finite subsets of $S_\gamma$
such that $|Y_n| \to \infty$ and the following relations hold
  \begin{equation}\label{hypothesis 1}
{1 \over |Y_n|} \sum_{y \in Y_n} f(y) \; \to \; \int f d\mu
  \end{equation}
for any bounded continuous function $f: \Gamma\to \C$,
  \begin{eqnarray}\label{hypothesis 2}
&& \sup_{n \in \N} {1 \over |Y_n|} \sup_{x \in Y_n} \sum_{y \in
Y_n \setminus \{x\} } G_k (x-y) < \alpha\,,
\\ \label{hypothesis 3}
&& \sup_{x \in Y_n} \left| {1 \over |Y_n|} \sum_{y \in Y_n
\setminus \{x\} } \sigma(y) G_k (x-y) - (R^k_{\D x,\mu} \sigma)(x)
\,\right| \; \to \; 0
  \end{eqnarray}
for $n \to \infty$. The operators $H_{Y_n,\alpha}$ and $H_{\gamma
m}$ defined in Section~\ref{prelim} then satisfy the relation
$H_{Y_n,\alpha} \to H_{\gamma m}$ in the strong resolvent sense as
$n\to\infty$ .
\end{theorem}
{\sl Proof:} Since both Hamiltonians $H_{Y_n,\alpha}$ and
$H_{\gamma m}$ are self-adjoint it is sufficient to prove the weak
convergence, i.e. to check that
$$ I_n = (\phi, (H_{Y_n,\alpha}-z)^{-1}\psi -(H_{\gamma
m}-z)^{-1}\psi )_{L^2(\R^2)} \; \to 0 \quad \mathrm{as} \quad n
\to \infty $$
holds for arbitrary $\psi,\phi \in L^2(\R^2)$. Using the formulae
(\ref{sing krein alt}) and (\ref{point krein alt}) for the
resolvents we arrive at
  \begin{eqnarray*}
I_n &=& \sum_{y' \in Y_n} {1\over |Y_n|} \,(R_0^k \bar{\phi})(y')
\left[ |Y_n| q_{y'} -\sigma(y') \right] \\ &&
+ {1\over |Y_n|} \sum_{y' \in Y_n} (R_0^k \bar{\phi})(y')\, \sigma(y')
- \int (R^k_{\mu,\D x} \bar{\phi})(y) \,\sigma(y) \mu (d y).
  \end{eqnarray*}
The sum of the last two terms tends to zero as $n \to \infty$,
which follows from the hypothesis (\ref{hypothesis 1}). Since the
function $R_0^k \bar{\phi}$ is continuous and bounded, as we have
already mentioned, it is enough to prove the following claim,
$$ {1\over |Y_n|}\, \| v^{(n)} \|_1 \to 0 \quad \mathrm{as} \quad
n \to \infty\ $$
for the $\ell^1$ norm, where $(v^{(n)})_y := |Y_n| q_y-\sigma(y)$,
$y\in Y_n$.

To this end, we employ the equations (\ref{eq sigma}) and (\ref{eq
q}) and we substitute the term $R^k_{\mu, \D x} \psi$ in one
equation from the other. The equation (\ref{eq sigma}) holds
$\mu$-a.e., so we must consider continuous representatives of the
functions involved here. In this way we get
  \begin{eqnarray}
\alpha\, \sigma(y) - (R_{\mu,\mu}^k \sigma) (y) &=& {1 \over 2\pi}
\left[ 2\pi \alpha(y)|Y_n| + \log\left( {i k\over 2} \right)
+C_E \right] q_y \nonumber \\
&& - \sum_{y'\in Y_n, y'\neq y} G_k(y-y') q_{y'}.
  \end{eqnarray}
By adding two extra terms to both sides of the equation we arrive
at
  \begin{eqnarray}
{1 \over |Y_n|} \sum_{y'\in Y_n}[\Lambda_{Y_n,\alpha}(k^2;y,y')
(q_{y'} |Y_n|-\sigma(y'))] = -{ \log( i k/2)+C_E \over 2\pi |Y_n|}
\,\sigma(y) \nonumber \\
+ {1 \over |Y_n|} \sum_{y'\in Y_n, y'\neq y} G_k(y-y') \sigma(y')
-\int G_k(y-y') \sigma(y') \mu(d y'). \nonumber
  \end{eqnarray}
We denote the vector on the r.h.s by $w^{(n)}$, then the previous formula
reads $(1/|Y_n|) \Lambda_{Y_n,\alpha}(z).v^{(n)}=w^{(n)}$.
Since we assume that the inequality (\ref{hypothesis 2}) holds, the
Lemma~\ref{lemma: lambda} is applicable here. Therefore there exists
$n_0 \in \N$ such that the matrix ${1 \over |Y_n|} \Lambda_{Y_n,\alpha}
(z)$ is invertible for all $n>n_0$. Then we can write
$$
{1 \over |Y_n|}\| v^{(n)} \|_1 \leq {1 \over |Y_n|}
\Big\| \left({1 \over |Y_n|} \Lambda_{Y_n,\alpha}(z) \right)^{-1}
\Big\|_{\infty,1} \| w^{(n)} \|_\infty.
$$
From the Lemma~\ref{lemma: lambda} and the relation $\| A
\|_{\infty,1} \leq |Y_n|\| A \|_{2,2}$, $A$ is a
operator acting on $\C^{|Y_n|}$, we conclude that the operator
norm in the inequality above is bounded by $|Y_n|C$ for some $C>0$.
Finally, let us look on the norm $\parallel w^{(n)} \parallel_\infty$:
the first term of $(w^{(n)})_y$ converges to zero uniformly w.r.t $y$
as $n \to \infty$ (recall that $\sigma$ is bounded) and the hypothesis
(\ref{hypothesis 3}) yields that remaining two terms go to zero
as well.
\QED


\setcounter{equation}{0}
\section{Soft ring graphs} \label{softring}

Let us now pass to examples. The first class of graphs to which we
apply the approximation developed in the Section \ref{main result}
are rings, both full or open. Since the spectral properties of
Hamiltonians with interaction supported by these ring graphs were
already explored in the paper \cite{ET} -- see also the
three-dimensional analogue discussed earlier in \cite{AGS} -- this
gives us an opportunity to compare the approximation with the
``exact'' results, in particular, to asses the convergence rate of
the approximation.

Consider a circle $\Gamma:=\{ x \in \R^2: |x|=R \}$ with the
radius $R>0$ and let $\gamma$ be a function from $\Gamma$ to $\R$
such that $\gamma(x)=\gamma \chi_{[0,2\pi-\theta]}(\varphi)$,
where $x=(R,\varphi)$, $\gamma>0$ and $0 \leq \theta < 2\pi$. The
Schr\"odinger operator with the $\delta$ interaction supported by
$\Gamma$ given formally by (\ref{formal}) will be denoted by
$H_{\gamma,R}$; it can be given meaning in either of the two ways
described in Section \ref{prelim}. The spectrum and eigenfunctions
are found easily in the full ring case, $\theta=0$, when
$H_{\gamma,R}$ is reduced by the angular momentum subspaces. Every
eigenstate except the ground state is twice degenerate. In
contrast, for a cut ring, $\theta>0$, the spectrum is simple; the
problem can be solved numerically by the mode-matching method
\cite{ET}.

\begin{figure}[!t]
\begin{center}
\includegraphics[height=6cm, width=9cm]{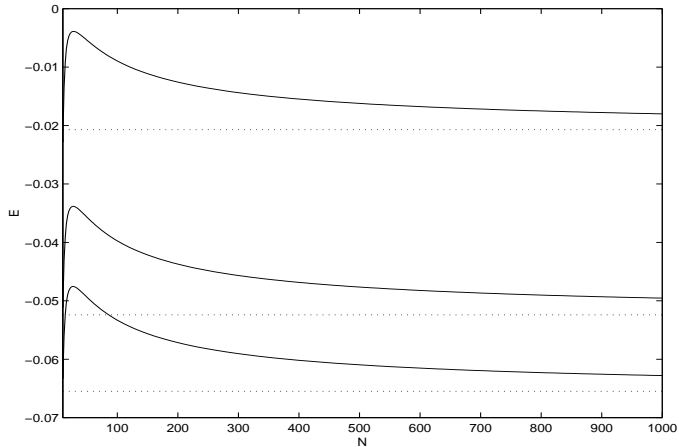}
\end{center}
\vspace{-0.5cm} \caption{The dependence of eigenvalues of
$H_{Y,\alpha}$ on the number of point potentials $N$ for
$\gamma=0.5$ and the ring graph with $R=10$. The dotted lines are
the exact eigenvalues $E_0=-0.0655$, $E_1=-0.0524$, and
$E_2=-0.0207$.} \label{fig:conv1}
\end{figure}

We start the presentation of the numerical results with the full
ring. First, we put $R=10$ and $\gamma=0.5$. The discrete spectrum
of $H_{\gamma,R}$ consists of three eigenvalues, the ground state
corresponds to the angular momentum $l=0$, the other two
correspond to $l=\pm1,\pm2$, respectively, and they are twice
degenerate; the number of eigenvalues is given be the inequality
$\gamma R>2|l|$. The choice of the approximating point-potential
operators $H_{Y_n,\alpha}$ is obvious -- $N$ point potentials are
spread periodically all over the circle and the coupling constant
$\alpha$ equals $N /(2\pi R\gamma)$. The task of finding the
eigenvalues $E$ of $H_{Y_n,\alpha}$ means to solve the implicit
equation
  \begin{equation}\label{point eigv}
\det \Lambda_{Y,\alpha}(E)=0.
  \end{equation}
We plot the eigenvalues of $H_{Y,\alpha}$ as $N=|Y|$ increases in
Fig.~\ref{fig:conv1}.

\begin{figure}[!t]
\begin{center}
\includegraphics[height=6cm, width=9cm]{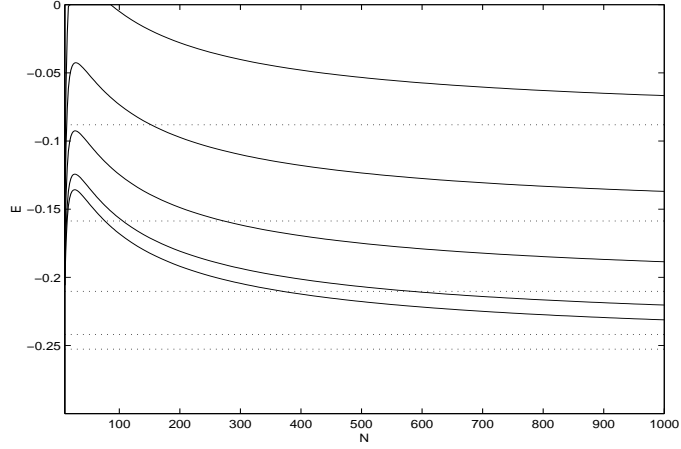}
\end{center}
\vspace{-0.5cm} \caption{The dependence of eigenvalues of
$H_{Y,\alpha}$ on the number of point potentials $N$ for
$\gamma=1$ and the ring graph with $R=10$ . The dotted lines are
the exact eigenvalues $E_0=-0.253$, $E_1=-0.243$, $E_2=-0.21$,
$E_3=-0.159$, and $E_4=-0.0881$.} \label{fig:conv2}
\end{figure}
\begin{figure}[!t]
\begin{center}
\includegraphics[height=6cm, width=9cm]{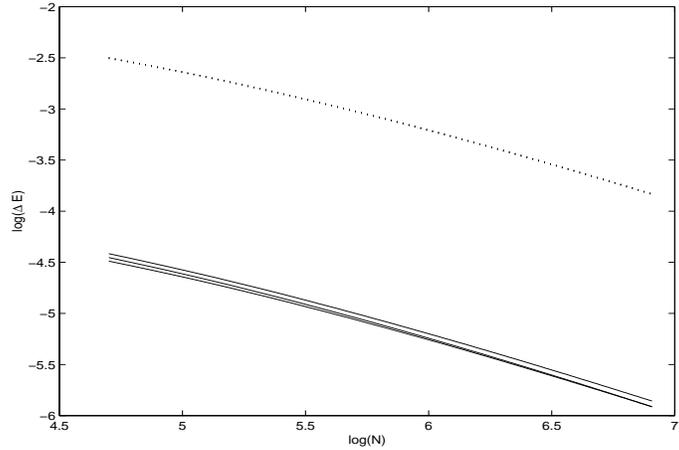}
\end{center}
\vspace{-0.5cm}
\caption{The dependence of the error on the number of
point potentials $N$ in the logarithmic scale.
Dotted line corresponds to $R=10$ and
$\gamma=0.5$ and the solid line corresponds to $R=10$ and
$\gamma=1$.} \label{fig:convdif}
\end{figure}

In case of a stronger interaction, $\gamma=1$, one gets a similar
picture, just the number of levels rises to five, see
Fig.~\ref{fig:conv2}. Note that the convergence of eigenvalues is
slower than it is for the previous system. To estimate the rate of
convergence, we calculate the difference between the exact
eigenvalue and the eigenvalue computed using the approximation.
The result is shown in Fig.~\ref{fig:convdif}, we observe that the
aforementioned difference decays roughly as $N^{-a}$ with $a$
being less than 1.

\begin{figure}[!t]
\begin{center}
\includegraphics[height=6cm, width=9cm]{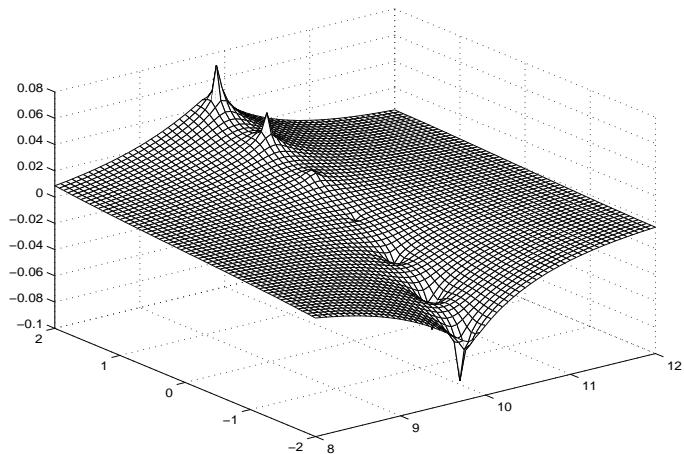}
\end{center}
\vspace{-0.5cm}
\caption{A detail of the wavefuction near the intersection
of the circle and one of the nodal lines.
It is the wavefunction of the third excited state
($l=3$) for $R=10$, $\gamma=5$ and $|Y|=100$.} \label{fig:cir-eigf1}
\end{figure}

\begin{figure}[!t]
\begin{center}
\includegraphics[height=6cm, width=9cm]{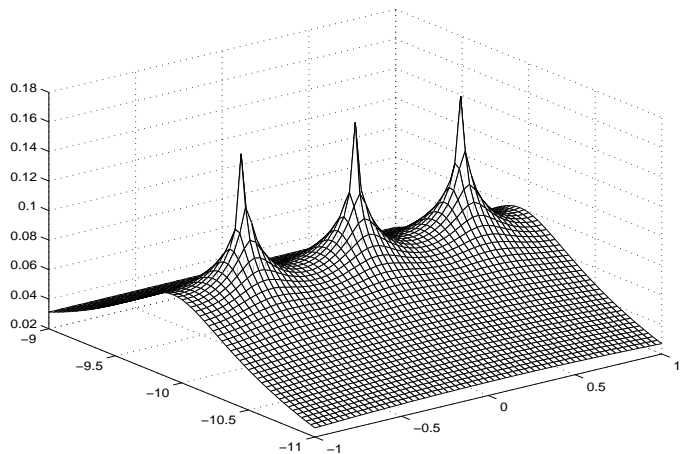}
\end{center}
\vspace{-0.5cm}
\caption{A detail of the wavefunction of the third excited state
($l=3$) for $R=10$, $\gamma=5$ and $|Y|=100$.} \label{fig:cir-eigf2}
\end{figure}

The approximation by point-potential Schr\"odinger operators
allows us to find easily the eigenfunctions. By \cite[Thm.~II.4.2]
{AGHH}, they can be written as a linear combination of the free
Green functions:
  \begin{equation}\label{point eigf}
\psi_0 (x) = \sum_{y \in Y} c_y G_{k_0}(x-y),
  \end{equation}
where $k_0^2$ is the eigenvalue, i.e. $\det \Lambda_{Y,\alpha}
(k_0^2)=0$, and $c$ is the solution to $\Lambda_{Y,\alpha}
(k_0^2)c=0$. The eigenfunctions obtained in this way behave as one
expects: they decrease exponentially if moving transversally away
of the circle, and for $l>0$ they copy a sine function if moving
along the circle. A closer inspection of the eigenfunctions
(\ref{point eigf}) shows, of course, a logarithmic peak at the
site of each point potential, as pictures Fig.~\ref{fig:cir-eigf1}
and Fig.~\ref{fig:cir-eigf2} demonstrate. To our opinion the
contributions to energy coming from these spikes are responsible
for the slow convergence of the approximation.

\begin{figure}[!t]
\begin{center}
\includegraphics[height=5cm, width=5cm]{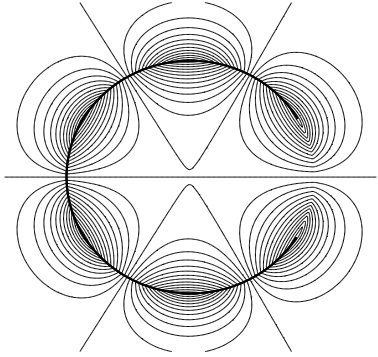}
\end{center}
\vspace{-0.5cm}
\caption{The wavefunction of the fifth excited state of
$H_{\gamma,R}$ for $R=10$, $\gamma=1$ and $\theta=\pi/3$.
The solid line represents the ring $\Gamma$.}
\label{fig:cir-e5}
\end{figure}

\begin{figure}[!t]
\begin{center}
\includegraphics[height=5cm, width=6cm]{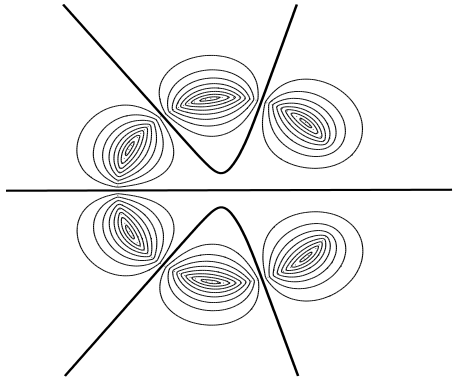}
\end{center}
\vspace{-0.5cm}
\caption{The wavefunction of the fifth excited state of
$H_{Y,\alpha}$ for $|Y|=1000$, which approximates the fifth
excited state of $H_{\gamma,R}$ from Fig.~\ref{fig:cir-e5}.
The solid lines represent the nodal lines, the dotted line represents
the graph $\Gamma$.}
\label{fig:cir-app5}
\end{figure}

Next, we consider an open ring, $\theta=\pi/3$, i.e. one sixth of
the perimeter is missing. For example, the fifth excited state for
$R=10$ and $\gamma=1$ has the energy $E_5=-0.151$, the
corresponding eigenfunction is shown in Fig.~\ref{fig:cir-e5}. The
approximation by 1000 point potentials yields the energy
$E'_5=-0.116$ and the corresponding eigenfunction shown in
Fig.~\ref{fig:cir-app5}.


\setcounter{equation}{0}
\section{Star-shaped graphs} \label{stargraph}

Another class of leaky-graph systems to which we will apply the
approximation by point-interaction Hamiltonians are the star-graph
Hamiltonians where $\Gamma$ is a collection of segments coupled at
a point. In distinction to the previous section no ``direct''
method to solve the problem is available in this case, and thus
the approximation represents the only way how to obtain a
numerical description of the eigenvalues and eigenfunctions.

First we want to make a remark about finiteness of such graphs.
Our interest concerns primarily infinite stars in which the arms
are halflines, in particular, since they support localized states
despite the fact that the graph geometry would allow escape to
infinity. On the other hand, the above approximation result
applies to finite graphs only because it requires $\int \gamma
m$ to be finite. Nevertheless, the result can be used, due to the
fact that the infinite star Hamiltonian is approximated, again in
the strong resolvent sense, by a family of operators with cut-off
stars. To justify this claim, it is sufficient to realize that the
corresponding family of quadratic forms is by (\ref{form})
monotonous and bounded from below, so that Thm.~VIII.3.11 of
\cite{K} applies.

Let us thus discuss in the beginning what can be derived
analytically about infinite star graphs. Given an integer $N\ge
2$, consider an $(N\!-\!1)$-tuple $\beta=\{ \beta_1,\dots,
\beta_{N-1}\}$ of positive numbers such that
$$ \beta_N := 2\pi - \sum_{j=1}^{N-1} \beta_j > 0\,.  $$
Denote $\vartheta_j:= \sum_{i=1}^j \beta_j$, where we put
conventionally $\vartheta_0=0$. Let $L_j$ be the radial halfline,
$L_j:= \{ x\in\R^2:\; \arg x=\vartheta_j \}$, which is naturally
parametrized by its arc length $s=|x|$. The support of
the interaction is given by $\Gamma \equiv
\Gamma(\beta) := \bigcup_{j=0}^{N-1} L_j$.

The star-graph Hamiltonians $H_N(\beta)$ are defined formally by
(\ref{formal}) as  Schr\"odinger operators with an attractive
potential supported by the graph $\Gamma$ and with the coupling
constant $\gamma$, and the proper meaning is given to this
operator in the way described in Section \ref{prelim}.

\begin{remark}
{\rm  Properties of the operator $H_N(\beta)$ certainly depend
on the order of the angles in $\beta$. However, the operators
obtained from each other by a {\em cyclic} permutation are
obviously unitarily equivalent by an appropriate rotation of the
plane.}
\end{remark}
Let us mention two trivial cases:
\begin{example}
{\rm  (a) $\,H_2(\pi)\,$ corresponding to a straight line can be
written as $h_{\gamma}\otimes I + I\otimes (-\partial_y^2)$, where
$h_{\gamma} = -\partial_x^2 -\gamma \delta(x)$ is the one-center
point-interaction Hamiltonian \cite{AGHH} on $L^2(\R)$.
Consequently, its spectrum is purely a.c. and equal to
$[-\gamma^2/4, \infty)$.
\\ [1mm]
(b) $\,H_4(\beta_s)\,$ with $\beta_s= \left\lbrace {\pi\over 2},
{\pi\over 2}, {\pi\over 2} \right\rbrace$ again allows separation
of variables being $h_{\gamma}\otimes I + I\otimes h_{\gamma}$.
Hence the a.c. part of $\sigma(H_4(\beta_s))$ is the same as
above, and in addition, there is a single isolated eigenvalue
$-\gamma^2/2$ corresponding to the eigenfunction $(2\gamma)^{-1}
\e^{-\gamma(|x|+|y|)/2}$. }
\end{example}

\subsection{The essential spectrum}

First we notice that the essential spectrum of $H_N(\beta)$ does
not extend below that of $H_2(\pi)$ corresponding to a straight
line.
\begin{proposition}
$\;\inf \sigma_\mathrm{ess}(H_N(\beta)) \ge -\,{\gamma^2\over
4}\:$ holds for any $N$ and $\beta$.
\end{proposition}
{\sl Proof:} By Neumann bracketing. We decompose the plane into a
finite union
\begin{equation} \label{Neub}
P\cup \left( \bigcup_j\: (S_j \cup W_j) \right)\,, \end{equation}
where $W_j$ is a wedge of angle $\beta_j$, $S_j$ is a halfstrip
centered at $L_j$ which is obtained by a Euclidean transformation
of $\R^+\times[\ell,\ell]$, and $P$ is the remaining polygon
containing the vertex of $\Gamma$. Imposing Neumann boundary
conditions at the common boundaries, we get a lower bound to
$H_N(\beta)$ by an operator which is a direct sum corresponding to
the decomposition (\ref{Neub}). The wedge parts have an a.c.
spectrum in $\R^+$ while the polygon has a purely discrete
spectrum. Finally, the halfstrip spectrum is a.c. again and
consists of the interval $[\epsilon(\ell),\infty)$, where
$\epsilon(\ell)$ is the lowest eigenvalues of $( -\partial_y^2 -\gamma
\delta(y))_N$ on $L^2([-\ell,\ell])$. It is straightforward to see
that to any $\eta < -\gamma^2/4$ there is $\ell$ such that
$\epsilon(\ell)>\eta$, and since the decomposition (\ref{Neub})
can be chosen with $\ell$ arbitrarily large, the proof is
finished. \quad \QED \vspace{3mm}

In fact, the essential spectrum is exactly that of a straight
line.
\begin{proposition}
$\;\sigma_\mathrm{ess}(H_N(\beta)) = [ -\gamma^2/4, \infty )\:$
holds for any $N$ and $\beta$.
\end{proposition}
{\sl Proof:} In view of the previous proposition, it is sufficient
to check that $\sigma_\mathrm{ess} (H_N(\beta)) \supset
[-\gamma^2/4,\infty)$. Given a function $\phi\in
C_0^{\infty}([0,\infty))$ with $\|\phi\|=1$ and $\phi(r)=1$ in the
vicinity of $r=0$, we construct
$$ \psi_n(x;p,x_n):= \frac{1}{n\sqrt{2\gamma}}
\:\phi\left(\frac{1}{n}|x-x_n|\right)\, \e^{-\gamma|x^{(2)}|/2}\,
\e^{ipx^{(1)}} $$
with $p\ge 0$ and $x=\left(x^{(1)}, x^{(2)} \right)$, where the
points $x_n$ can be chosen, e.g., as $(n^2,0)$. It is easy to see
that the vectors $\psi_n\to 0$ weakly as $n\to\infty$ and that
they form a Weyl sequence of $H_N(\beta)$ referring to the value
$-\gamma^2/4+p^2$. This yields the desired result. \quad \QED
\vspace{3mm}


\subsection{The discrete spectrum}

The first question concerns naturally the existence of isolated
eigenvalues. It follows from two observations of which one is
rather simple.
\begin{proposition} \label{addleg}
$\;H_N(\beta) \ge H_{N+1}(\tilde\beta)\:$ holds for any $N$ and
angle sequence $\tilde\beta = \{ \beta_1, \dots, \beta_{j-1},
\tilde\beta_j^{(1)}, \tilde\beta_j^{(2)}, \beta_{j+1}, \dots,
\beta_{N-1}\}$ with $\tilde\beta_j^{(1)} + \tilde\beta_j^{(2)} =
\beta_j$.
\end{proposition}
{\sl Proof} follows directly from the definition by the quadratic
form (\ref{form}).
\quad \QED \vspace{3mm}

\noindent On the other hand, the second one is rather nontrivial.
It has been proven in \cite{EI} for a wide class of piecewise
continuous non-straight curves which includes, in particular, a
broken line.
\begin{proposition} \label{exist-pr}
$\;\sigma_\mathrm{disc} (H_2(\beta))\:$ is nonempty unless
$\beta=\pi$.
\end{proposition}
Combining these two results with the minimax principle (recall
that $H_N(\beta)$ is below bounded) we arrive at the following
conclusion.
\begin{theorem} \label{exist-thm}
$\;\sigma_\mathrm{disc} (H_N(\beta))\:$ is nonempty except if
$N=2$ and $\beta=\pi$.
\end{theorem}

Next one has to ask how many bound states does a star graph
support. The answer depends on its geometry. There are situations,
however, where their number can be large.
\begin{theorem} \label{card-thm}
Fix $N$ and a positive integer $n$. If at least one of the angles
$\beta_j$ is small enough, $\mathrm{card\,}\left(
\sigma_\mathrm{disc} (H_N(\beta)) \right) \ge n$.
\end{theorem}
{\sl Proof:} In view of Proposition~\ref{addleg} it is again
sufficient to check the claim for the operator $H_2(\beta)$.
Choose the coordinate system in such a way that the two ``arms''
correspond to $\arg\theta = \pm\beta/2$. We employ the following
family of trial functions
\begin{equation} \label{trial}
\Phi(x,y) = f(x)g(y)
\end{equation}
supported in the strip $L\le x \le 2L$, with  $f\in C^2$
satisfying $f(L)=f(2L)=0$, and
$$ g(y) = \left\{ \begin{array}{cll} 1 & \quad \dots \quad &
|y|\le 2d \\ \e^{-\gamma(|y|-2d)} & \quad \dots \quad & |y|\ge 2d
\end{array} \right.
$$
with $d:= L \tan(\beta/2)$. Let us ask under which conditions the
value of the shifted energy form
$$ q[\Phi] := \|\nabla\Phi\|^2 - {2\gamma\over \cos(\beta/2)}
\|f\|^2 + {\gamma^2\over 4} \|\Phi\|^2 $$
is negative. Since $\|g\|^2 =4d+\gamma^{-1}$ and $\|g'\|^2
=\gamma$, this is equivalent to
$$ {\|f'\|^2\over \|f\|^2} < \gamma^2\: {2\sec{\beta\over 2} -
\gamma d - {5\over 4} \over 1+4\gamma d }\,. $$
By minimax principle the system there will be at least $n$
isolated eigenvalues provided
$$ \left( {\pi n\over d}\, \tan{\beta\over 2} \right)^2 =\:
\inf_{M_n^\perp}\, \sup_{M_n}\, {\|f'\|^2\over \|f\|^2}\: <
\gamma^2\: {2\sec{\beta\over 2} - \gamma d - {5\over 4} \over
1+4\gamma d }\,, $$
where $M_n$ means an $n$-dimensional subspace in $L^2([L,2L])$,
i.e., if
$$ n^2 < {\gamma^2 d^2\over \pi^2}\: \left(\cot{\beta\over
2}\right)^2 \, {2\sec{\beta\over 2} - \gamma d - {5\over 4} \over
1+4\gamma d }\,. $$
Now one should optimize the r.h.s. w.r.t. $\gamma d$, but for a
rough estimate it is sufficient to take a particular value, say
$\gamma d= \sec{\beta\over 2} - {5\over 8}$ which yields
\begin{equation} \label{n-est}
n < {1\over 16\pi}\: \cot{\beta\over 2} \: { \left(
8\sec{\beta\over 2} - 5 \right)^{3/2} \over \left(
8\sec{\beta\over 2} - 3 \right)^{1/2}}\;;
\end{equation}
it is obvious that the last inequality is for any fixed $n$
satisfied if $\beta$ is chosen small enough. \quad \QED
\vspace{3mm}

\begin{corollary} \label{card-cor}
Independently on $\beta$, $\:\mathrm{card\,}\left(
\sigma_\mathrm{disc} (H_N(\beta)) \right)$ exceeds any fixed
integer for $N$ large enough.
\end{corollary}
\begin{remark}
{\rm  The estimate used in the proof also shows that the number of the bound
states for a sharply broken line is roughly proportional to the inverse angle,
$$ n \ageq {3^{3/2}\over 8\pi \sqrt{5}}\: \beta^{-1} $$
as $\beta\to 0$. This is the expected result, since the number is
given by the length of the effective potential well which exists
in the region where the two lines are so close that they roughly
double the depth of the transverse well. }
\end{remark}


\subsection{The Birman-Schwinger approach}

Now we are going to show how the spectral problem for the
operators $H_N(\beta)$ can be reformulated in terms of suitable
integral operators. We will employ the resolvent formula for
measure perturbations of the Laplacian derived in \cite{BEKS}.
Notice that this technique was used in \cite{EI} to derive a
result which implies our Proposition~\ref{exist-pr}.

Since the operators $H_N(\beta)$ are defined by the quadratic form
(\ref{form}) they satisfy the generalized Birman-Schwinger
principle. If $k^2$ belongs to the resolvent set of $H_N(\beta)$
we put $R^k_{\gamma,\Gamma} := (H_N(\beta)-k^2)^{-1}$.
We already know the Krein-like formula for the resolvent from
Thm~\ref{gen krein}, here it has the form
$$ R^k_{\gamma,\Gamma} = R_0^k + \gamma R^k_{\D x,m} [I-\gamma R^k_{m,m}]^{-1}
R^k_{m,\D x} ,$$
with $m$ denoting the Dirac measure on $\Gamma$.
One can express the generalized BS
principle as follows \cite{BEKS}:
\begin{proposition} \label{BS}
$\:\dim\ker(H_N(\beta)-k^2) = \dim\ker(I-\gamma R^k_{m,m})$
for any $k$ with $\im k>0$.
\end{proposition}

Consequently, the original spectral problem is in this way
equivalent to finding solutions to the equation
\begin{equation} \label{speceq}
\RR^{\kappa}_{\gamma,\gamma} \phi=\phi
\end{equation}
in $L^2(\Gamma)$, where $\RR^{\kappa}_{\gamma,\Gamma}:= \gamma
R^{i\kappa}_{m,m}$. Furthermore, in analogy with
\cite[Sec.~II.1]{AGHH} one expects that a non-normalized one
corresponding to a solution $\phi$ of the above equation is
$$ \psi(x) = \int_{\Gamma} G_{i\kappa}(x\!-\!x(s)) \phi(s)\, \D
s\,; $$
a proof of this claim can be found in the sequel to the paper
\cite{Po}.

\subsection{Application to star graphs}

Let us look now how the equation (\ref{speceq}) looks like for
graphs of the particular form considered here. Define
\begin{equation} \label{dist}
d_{ij}(s,s') \equiv d_{ij}^{\beta}(s,s') = \sqrt{s^2\! +{s'}^2\!
-2ss' \cos|\vartheta_j -\vartheta_i|}
\end{equation}
with $\vartheta_j -\vartheta_i = \sum_{l=i+1}^j \beta_l$, in
particular, $d_{ii}(s,s')= |s-s'|$. By $\RR^{\kappa}_{ij}(\beta) =
\RR^{\kappa}_{ji}(\beta)$ we denote the operator $L^2(\R^+) \to
L^2(\R^+)$ with the kernel
$$ \RR^{\kappa}_{ij}(s,s';\beta) := {\gamma\over 2\kappa}\: K_0
\left(\kappa d_{ij}(s,s') \right)\;; $$
then (\ref{speceq}) is equivalent to the matrix integral-operator
equation
\begin{equation} \label{speceq1}
\sum_{j=1}^N \left(\RR^{\kappa}_{ij}(\beta) - \delta_{ij}I \right)
\phi_j = 0\,, \quad i=1,\dots,N\,,
\end{equation}
on $\bigoplus_{j=1}^N L^2(\R^+)$. Notice that the above kernel has
a monotonicity property,
\begin{equation} \label{monot}
\RR^{\kappa}_{ij}(\beta) > \RR^{\kappa}_{ij}(\beta')
\end{equation}
if $|\vartheta_j -\vartheta_i| < |\vartheta'_j -\vartheta'_i|$.
This has the following easy consequence:
\begin{proposition} \label{mono2}
Each isolated eigenvalue $\epsilon_n(\beta)$ of $H_2(\beta)$ is an
increasing function of $\beta$ in $(0,\pi)$.
\end{proposition}
{\sl Proof:} Notice first that the eigenfunction related to
$\epsilon_n(\beta)$ is even with respect to the interchange of the
two halflines. Without loss of generality we may assume that $\arg
L_j= (-1)^{j-1}\beta/2$. The odd part of $H_2(\beta)$ then
corresponds to the Dirichlet condition at $x=0$. The spectrum of
this operator remains the same if we change $\arg L_1$ to
$\pi-\beta/2$. Removing then the Dirichlet condition, we get the
operator $H_2(\pi)$ with $\inf \sigma(H_2(\pi)) = \inf
\sigma_\mathrm{ess}(H_2(\beta))$, so by minimax principle no
eigenfunction of $H_2(\beta)$ can be odd.

On the symmetric subspace the diagonal matrix element of the
matrix integral operator $ \RR^{\kappa}(\beta)$ equals
$$ (\phi, \RR^{\kappa}(\beta)\phi) = 2(\phi_1,
\RR_{11}^{\kappa}\phi_1) + 2(\phi_1,
\RR_{12}^{\kappa}(\beta)\phi_1) $$
with the first term at the r.h.s. independent of $\beta$. Next we we
notice that $\{ \RR^{\kappa}(\beta) \}$ is a type (A) analytic
family around any $\beta\in(0,\pi)$; the derivative
${\D\over\D\beta} \RR^{\kappa}(\beta)$ is a bounded operator or
the form $\left( \begin{array}{cc} 0 & \DD_\beta \\ \DD_\beta & 0
\end{array} \right)$ where $\DD_\beta$ has the kernel
$$ \DD_\beta(s,s') = -\,{\gamma\over 2}\: K_1\left(\kappa
d^{\beta}_{12}(s,s') \right)\, {ss' \sin\beta \over
d^{\beta}_{12}(s,s')}<0\,. $$
At the same time $\{ \RR^{\kappa}(\beta) \}$ is a type (A)
analytic family w.r.t. $\kappa$ around any $\kappa\in(0,\infty)$
and the corresponding derivative is a bounded operator $\left(
\begin{array}{cc} \RR'_{11} & \RR'_{12} \\ \RR'_{12} & \RR'_{22}
\end{array} \right)$ with the kernel
$$ \RR'_{ij}(s,s') = -\,{\gamma\over 2\kappa^2}\, \left\lbrack
K_0\left(\kappa d_{ij}^{\beta}(s,s') \right) + \kappa
d_{ij}^{\beta}(s,s') K_1\left(\kappa d_{ij}^{\beta}(s,s') \right)
\right\rbrack < 0\,. $$
Let $\phi^{\beta}= {\phi_1^{\beta} \choose \phi_1^{\beta}}$ be a
normalized eigenvector of $\RR^{\kappa}(\beta)$ corresponding to
the eigenvalue $\lambda(\kappa, \beta) = (\phi^{\beta},
\RR^{\kappa}(\beta)\phi^{\beta})$. By Feynman-Hellmann theorem we
find
$$ {\D\over\D\beta}\, \lambda(\kappa, \beta) =
2\left(\phi_1^{\beta}, \DD_\beta \phi_1^{\beta}\right) < 0\,, $$
and similarly ${\D\over\D\kappa}\, \lambda(\kappa, \beta) < 0$.
The solution $\kappa=\kappa(\beta)$ of the implicit equation
$\lambda(\kappa, \beta)=1$ thus satisfies
$$ {\D\over\D\beta}\,\kappa(\beta) = -\, {\D\lambda/\D\beta \over
\D\lambda/\D\kappa} < 0\,, $$
and consequently, the eigenvalue $-\kappa(\beta)^2$ is increasing
w.r.t. $\beta$. \quad \QED


\subsection{Numerical results}

Having explained analytically how does the discrete spectrum of
$H_N(\beta)$ depend on the number of arms and the angles $\beta$,
we employ now the approximation by point-potential Schr\"odinger
operators to obtain the numerical results for cut-off stars which
would illustrate the above conclusions.

Consider a two-arms star graph $\Gamma_1$ with both arms having
the same length, $L_1=L_2=300$, and put $\gamma=0.1$. As we
already know, the threshold of the continuous spectrum of
$H_2(\beta)$ is $-\gamma^2/4=-0.0025$. We approximate $H_2(\beta)$
by point-potential Schr\"odinger operator $H_{Y,\alpha}$ that has
1 potential on the center of the star graph and $200$ equidistant
point potentials on each arm. Only the lower part of the discrete
spectrum of $H_{Y,\alpha}$ approximates the discrete spectrum of
the star-graph operator, while the upper part corresponds to the
interval $[-\gamma^2/4,0] \subset \sigma_{{\rm ess}}(H_2(\beta))$.
Therefore only the states with energy below the threshold may be
indeed understood as the approximation of the bound states of
$H_2(\beta)$.

\begin{figure}[!t]
\begin{center}
\includegraphics[height=6cm, width=9cm]{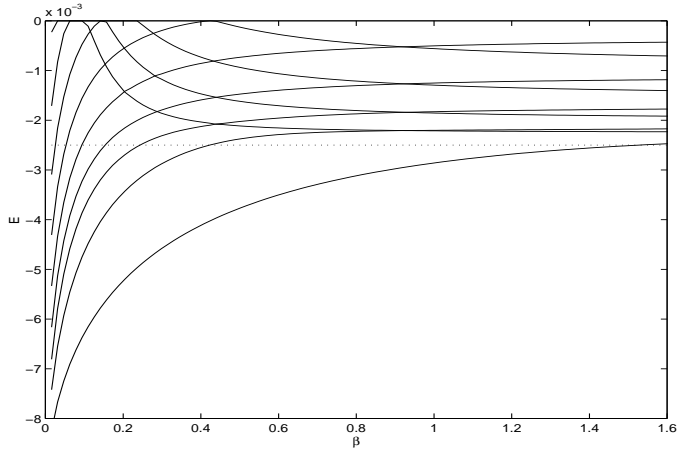}
\end{center}
\vspace{-0.5cm}
\caption{The dependence of the discrete spectrum of $H_{Y,\alpha}$
on the angle $\beta$ for the symmetric two-arms star $\Gamma_1$.
The dotted line represents the threshold $-\gamma^2/4$.}
\label{fig:star-spec1}
\end{figure}
\begin{figure}[!b]
\begin{center}
\includegraphics[height=6cm, width=9cm]{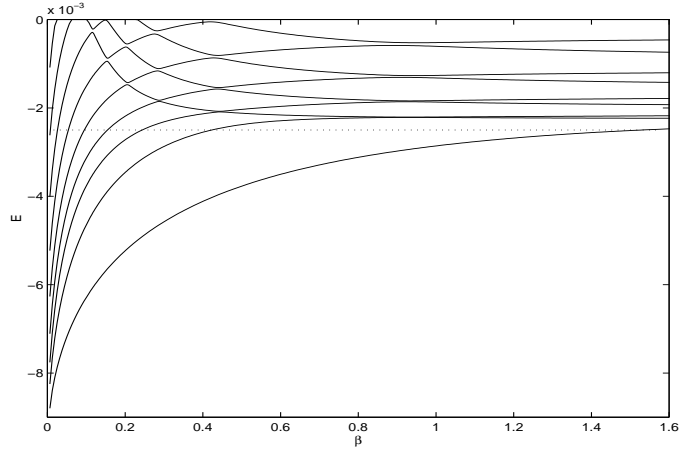}
\end{center}
\vspace{-0.5cm}
\caption{The dependence of the discrete spectrum of $H_{Y,\alpha}$
on the angle $\beta$ for the non-symmetric two-arms star $\Gamma_2$.
The dotted line represents the threshold $-\gamma^2/4$.}
\label{fig:star-spec2}
\end{figure}
The main result of the analytic argument presented above was the
dependence of the eigenvalues on the angle $\beta$: if $\beta$
decreases, the eigenvalues decrease and their number grows. The
discrete spectrum of $H_{Y,\alpha}$ for $\beta$ varying is in a
good agreement with this fact as Fig.~\ref{fig:star-spec1}
illustrates. The eigenvalue crossings we observe are actual, which
is a consequence of the symmetry of the graph $\Gamma_1$. For a
non-symmetric graph the crossings become avoided. To demonstrate
it in Fig.~\ref{fig:star-spec2}, we slightly change the length of
one arm, $L_2=306$, while the other parameters are preserved. Also
the setting of the point potentials approximating the new graph
$\Gamma_2$ stays the same, up to extra four potentials added on
the longer arm.

\begin{figure}[!b]
\begin{center}
\includegraphics[height=5cm, width=5cm]{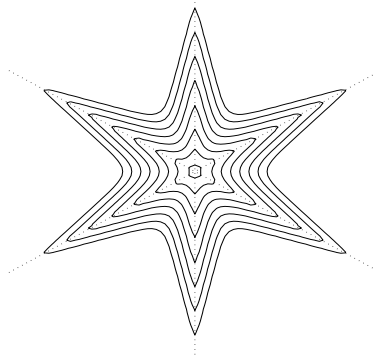}
\end{center}
\vspace{-0.5cm} \caption{The wavefunction of the ground state of
$H_{Y,\alpha}$, $E_0=-0.612$, which approximates the ground state
of $H_6(\beta)$ with $\beta$ being the 5-tuple of $\pi/3$ and
$\gamma=1$. The contours correspond to logarithmically scaled
horizontal cuts, the dotted lines represents the graph $\Gamma$.}
\label{fig:star-eigf0}
\end{figure}
\begin{figure}[!t]
\begin{center}
\includegraphics[height=5cm, width=6cm]{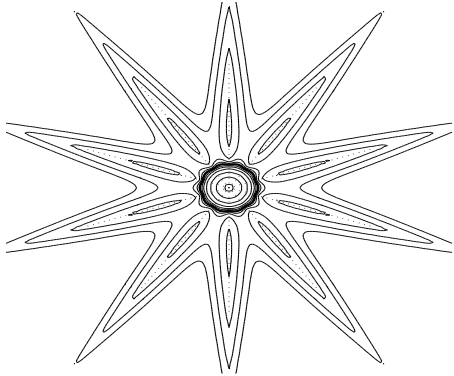}
\end{center}
\vspace{-0.5cm} \caption{The wavefunction of the third exited
state of $H_{Y,\alpha}$, $E_3=-0.265$. It approximates the third
excited state of $H_{10}(\beta)$ with $\beta$ being the 9-tuple of
$\pi/5$ and $\gamma=1$. The convention is the same as above, the
solid line is the nodal line.} \label{fig:star-eigf3}
\end{figure}
The expression (\ref{point eigf}) again yields the eigenfunctions
in the form of a sum of the free Green functions. We limit
ourselves to a pair of examples: the ground state of $H_6(\beta)$
in Fig.~\ref{fig:star-eigf0} and the third excited state of
$H_{10}(\beta)$ in Fig.~\ref{fig:star-eigf3}. The length of the
cut arms is 30, in the former case the approximating operator
$H_{Y,\alpha}$ has 601 point potentials and in the latter case the
number of point potentials is 1001. We see, in particular, that
for $N$ large enough $H_{N}(\beta)$ may have closed nodal lines;
it is an interesting question what is the minimal $N$ for which
this happens.


\setcounter{equation}{0}
\section{$L^2$-approach to resonances} \label{reson}

Our last example, as indicated in the introduction, concerns the
situation when $\Gamma$ is a single infinite curve; we want to see
whether resonances in the scattering of states propagating along
$\Gamma$ may be detected by inspecting the spectrum of the cut-off
problem with the curve of a finite length which is a parameter to
be varied.

\begin{figure}[!b]
\begin{center}
\includegraphics[height=4cm, width=6cm]{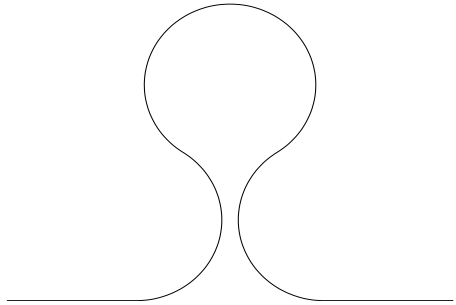}
\end{center}
\vspace{-0.5cm}
\caption{An example of the curve $\Gamma$: $R=10$ and $\Delta=1.9$.}
\label{fig:res-graph}
\end{figure}
\begin{figure}[!t]
\begin{center}
\includegraphics[height=7cm, width=10cm]{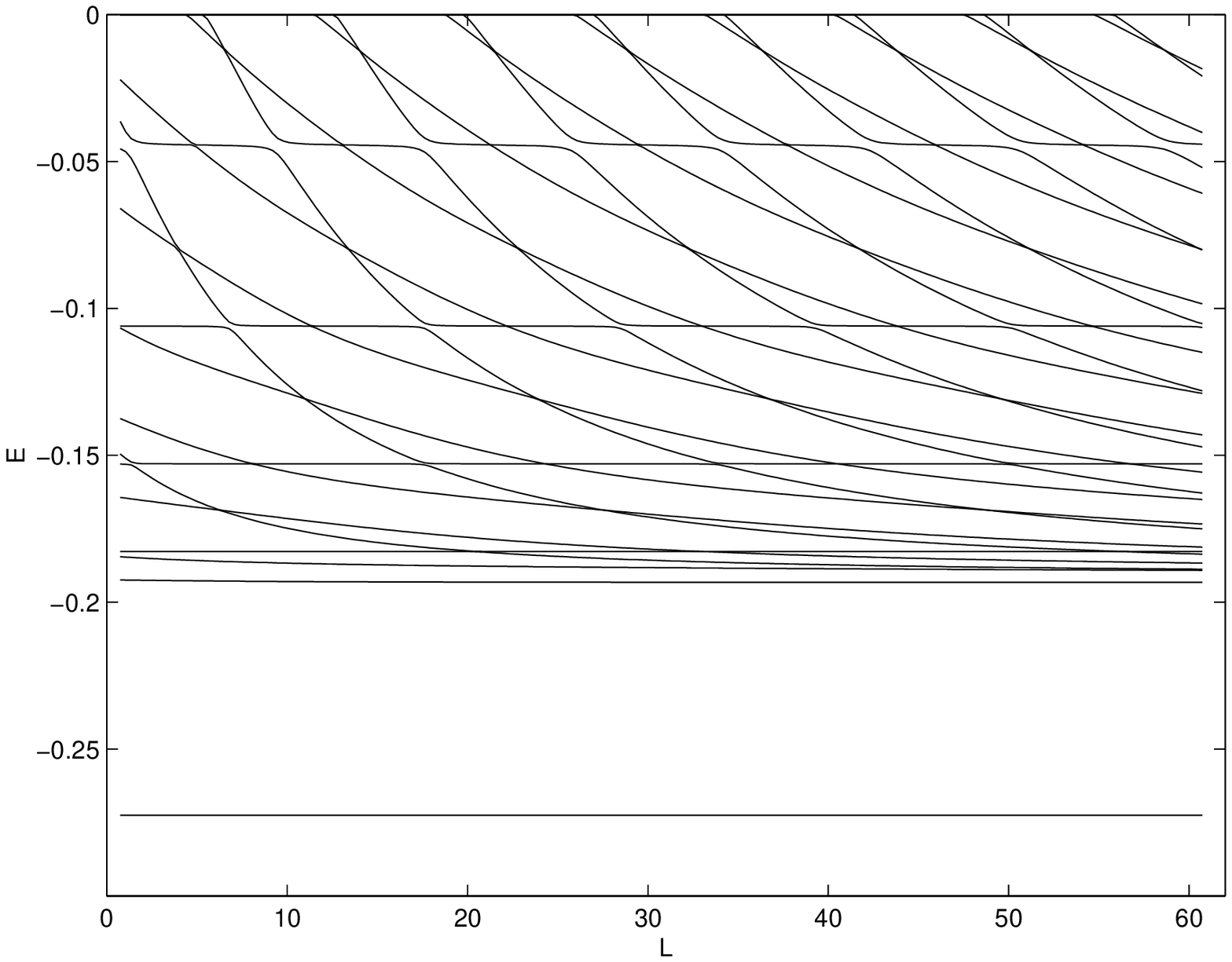}
\end{center}
\vspace{-0.5cm} \caption{The dependence of the discrete spectrum
of the $H_{Y,\alpha}$ on the length $L$ for $R=10$, $\gamma=1$ and
$\Delta=1.9$.} \label{fig:res-cir1}
\end{figure}
\begin{figure}[!t]
\begin{center}
\includegraphics[height=7cm, width=10cm]{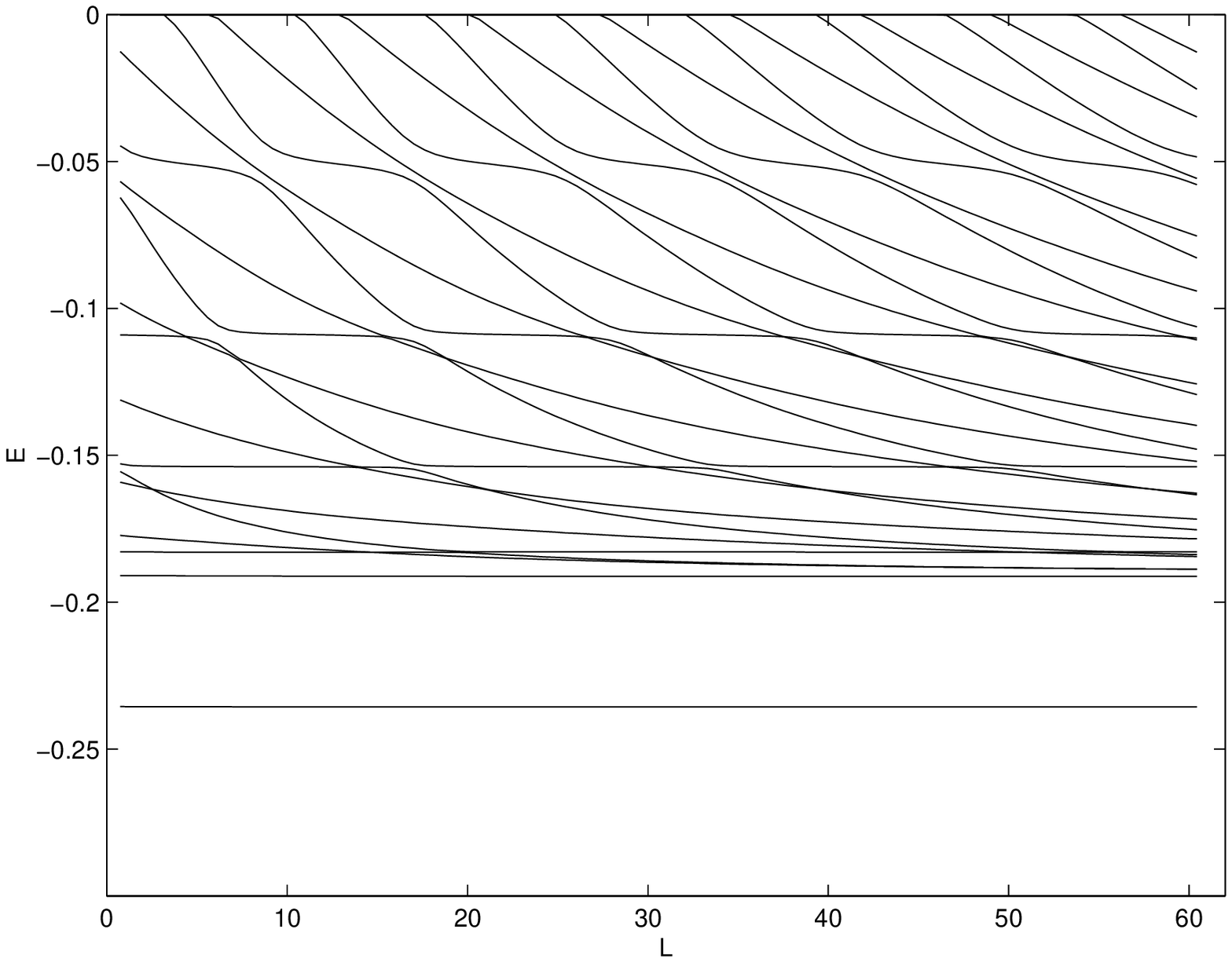}
\end{center}
\vspace{-0.5cm}
\caption{The dependance of the discrete spectrum of
$H_{Y,\alpha}$ on the length $L$ for $R=10$, $\gamma=1$ and
$\Delta=2.9$.}
\label{fig:res-cir2}
\end{figure}
\begin{figure}[!t]
\begin{center}
\includegraphics[height=7cm, width=10cm]{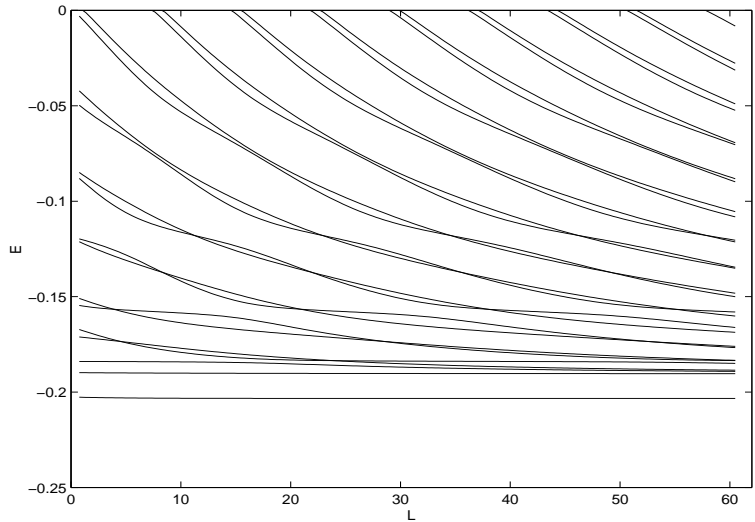}
\end{center}
\vspace{-0.5cm} \caption{The dependence of the discrete spectrum
of $H_{Y,\alpha}$ on the length $L$ for $R=10$, $\gamma=1$ and
$\Delta=5.2$.} \label{fig:res-cir3}
\end{figure}

Consider a curve $\Gamma$ from Fig.~\ref{fig:res-graph}: its
central part consists of three segments of a circle with the
radius $R=10$ and it has two infinite ``legs''. The distance
between two closest points of the curve, i.e. the bottleneck
width, is denoted by $\Delta$. The coupling constant $\gamma$
equals 1. We cut the ``legs'' of the graph to a finite length $L$
and we plot the eigenvalues computed using the approximation for
$L$ varying. The number of the point potentials involved in
$H_{Y,\alpha}$ is chosen so that the distance between every two
adjacent points equals $0.3$. For small values $\Delta=1.9$ and
$\Delta=2.9$ the tunneling effect occurs and we can see the
\emph{plateaux} in Figs.~\ref{fig:res-cir1}, \ref{fig:res-cir2},
which indicate the existence of resonances; the width of the
avoided crossings increase with $\Delta$ as expected. On the other
hand, for more open curve, $\Delta=5.2$, we get a different
picture, where the \emph{plateaux} are absent, see
Fig.~\ref{fig:res-cir3}.

\begin{figure}[!t]
\begin{center}
\includegraphics[height=7cm, width=10cm]{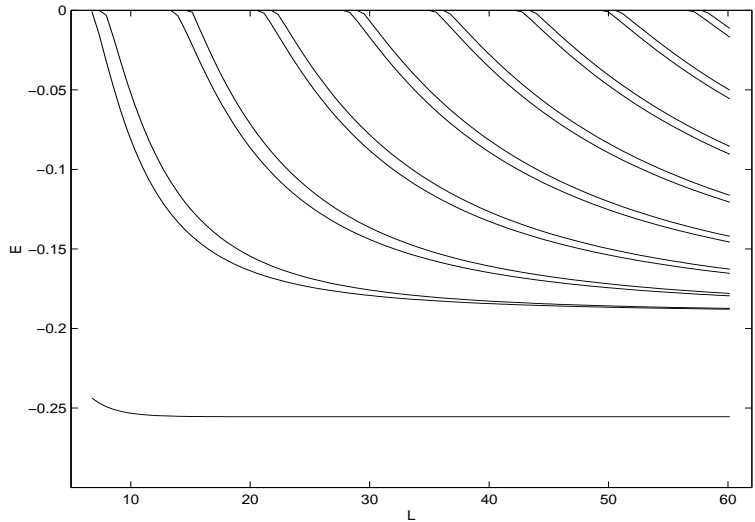}
\end{center}
\vspace{-0.5cm} \caption{The dependence of the discrete spectrum
of $H_{Y,\alpha}$ on the arm length $L$ for the angle
$\beta=\pi/4$ and $\gamma=1$.} \label{fig:res-star}
\end{figure}

The second graph type we are interested in are simple bends, in
terms of Section \ref{stargraph} it is a two-arms star $\Gamma
(\beta)$. As we have already mentioned, the avoided eigenvalue
crossings are not expected here because the transport along the
graph arm involves a single transverse mode -- this is confirmed
in Fig.~\ref{fig:res-star}, which shows the results of the cut-off
method for $\beta=\pi/4$.

\begin{figure}[!t]
\begin{center}
\includegraphics[height=7cm, width=10cm]{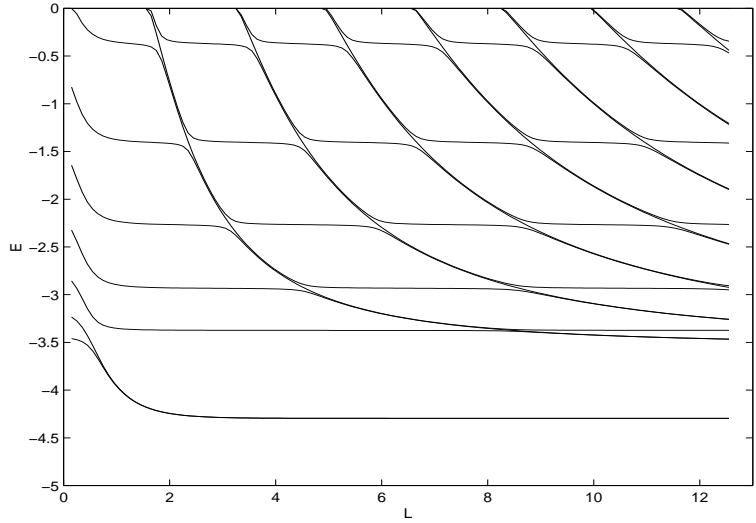}
\end{center}
\vspace{-0.5cm} \caption{The dependence of the discrete spectrum
of $H_{Y,\alpha}$ on the arm length $L$ for $R=10$,
$\theta=0.32\pi$ and $\gamma=5$.} \label{fig:res-stair1}
\end{figure}

\begin{figure}[!t]
\begin{center}
\includegraphics[height=7cm, width=10cm]{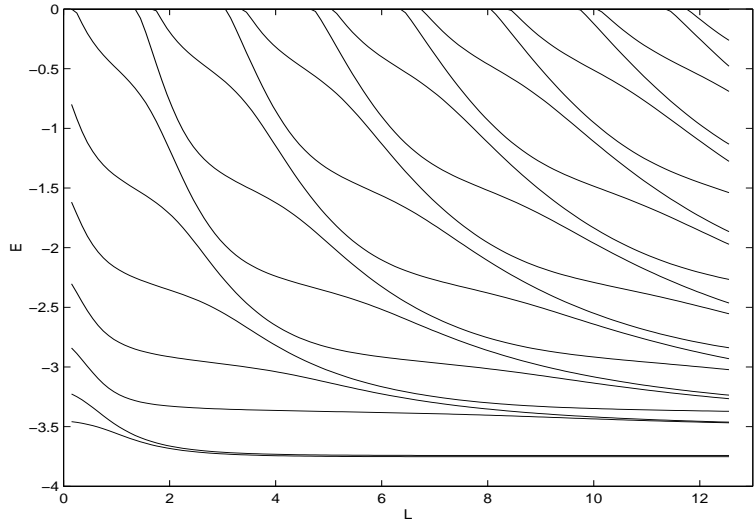}
\end{center}
\vspace{-0.5cm} \caption{The dependence of the discrete spectrum
of $H_{Y,\alpha}$ on the arm length $L$ for $R=10$,
$\beta=\pi$ and $\gamma=5$.} \label{fig:res-stair2}
\end{figure}

Finally to illustrate that resonances may also result from
multiple reflections rather than from a tunneling, we apply the
cut-off method to graphs which are of a stair-type, or Z-shaped.
They consists of three line segments: the central one has a finite
length $R$, i.e. the ``heigth'' of the stair, the other two are
parallel (cut-off) halflines. If the angle $\theta$ between the
segments is less than $\pi/2$ the term Z-shaped is more
appropriate. The results for $R=10$ and $\gamma=5$ are plotted in
Figs.~\ref{fig:res-stair1} and \ref{fig:res-stair2}. In the former
case the stair is ``skewed'' to $\theta=0.32\pi$, in the latter we
have $\theta=\pi$; the distance between the point potentials of
the approximating operator $H_{Y,\alpha}$ equals $0.1$. We find
avoided crossings, however, they are very narrow for
$\theta=0.32\pi$ and barely visible in the right-angle case; this
observation can be naturally understood in terms of the reflection
probability through a single bend -- compare with the angle
dependence of the spectrum on Fig.~\ref{fig:star-spec1}.



\subsection*{Acknowledgement}

The research was partially supported by the GAAS grant A1048101.
We appreciate the assistance of M.~O\v{z}ana in performing the numerical
computations, and of M.~Tater who supplied Fig.~\ref{fig:cir-e5}.


\bibliographystyle{plain}

\end{document}